\documentclass[preprint,12pt]{article}
\usepackage{subcaption}
\usepackage{graphicx}
\usepackage[round,authoryear]{natbib}
\usepackage{amssymb}
\usepackage{amsmath}
\newtheorem{theorem}{Theorem}
\newtheorem{lemma}{Lemma}
\newtheorem{definition}{Definition}

\newtheorem{prop}{Proposition}
\newcommand{\finish}{\hfill$\Box$\vspace{0.2cm}}

\newcommand{\prf}{\noindent{\bf Proof:\ }}

\providecommand{\keywords}[1]{\textbf{Keywords} #1}

\begin{document}

\title{A Generalized Correlated Random Walk Converging to Fractional Brownian Motion}



\author{Buket Co\c{s}kun \and Ceren Vardar-Acar \and Hakan Demirta\c{s}}



\maketitle

\begin{abstract}  We propose a new algorithm to generate a fractional Brownian motion, with a given Hurst parameter, $ H\in [1/2,1],$ using the correlated Bernoulli random variables with parameter $p,$ having a certain density. This density is constructed using the link between the correlation of multivariate Gaussian random variables and the correlation of their dichotomized binary variables and the relation between the correlation coefficient and the persistence parameter. We have proven that the normalized sum of trajectories of this proposed random walk yields a Gaussian process whose scaling limit is the desired fractional Brownian motion with the given Hurst parameter, $H\in [1/2,1].$
\end{abstract}	
\keywords{Correlated Random Walk, \and Dichotomized Binary Variables, \and Fractional Brownian motion, \and Gaussian Process}

\section{Introduction}\label{Section1}

Most of the real data displaying long-range dependence can be modeled using self-similar processes. Fractional Brownian motion (fBm) is one of the simplest models demonstrating long-range dependence. That is why this phenomenon has been widely used in many research areas and disciplines in recent years. For instance, in communication systems, \cite{leland1994self} used the increments of fBm for modeling Ethernet local area network (LAN) traffic. In the field of mathematical finance, \cite{rogers1997arbitrage} proposed an fBm model to explain the movement of share prices. In biology, \cite{lim2001fractional} utilized the discrete-time version of fBm to model the non-coding sequence of human DNA by recognizing a DNA sequence as a fractal random walk. Additionally, there is still an increasing interest in the use of fBm in many fields since it well captures the dependence behavior of the data. For this reason, its theoretical properties and path behaviors have been well studied and the interest in these studies still continues.
\begin{definition}
	Let $H$ be a constant belonging to $(0,1)$. A fractional Brownian motion (fBm), $B^{H}(t)_{t\geq0},$ is continuous, centered Gaussian process with Hurst parameter $H,$ and with covariance function;
	\begin{equation}
	E[B^{H}(t)B^{H}(s)]=\frac{1}{2}(t^{2H}+s^{2H}-|t-s|^{2H}).
	\end{equation}
	
\end{definition}
The fractional Brownian motion satisfies the following properties;
\begin{enumerate}
	\item[i.]$B^{H}(0)=0$, and $E(B^{H}(t))=0$  for all $t\geq 0.$
	\item[ii.] $B^{H}(t)$ has stationary increments that is $B^{H}(t+s)-B^{H}(s)$ has the same distribution with $B^{H}(t)$ for $s,t\geq 0$.
	\item[iii.] $B^{H}(t)$ is a Gaussian process.
	\item[iv.] The variance of $B^{H}(t)$ equals $t^{2H}$ for all $t\geq0$ and $H\in(0,1)$.
	\item[v.] $B^{H}(t)$ has continuous trajectories.
	\item[vi.] The increment process of fBm, $\{B^{H}(n+1)-B^{H}(n): n=0,1,2,...\},$ also called the fractional Gaussian noise(fGn) which are jointly Gaussian variates with zero mean and autocovariance function 
	\begin{eqnarray}
	\label{eq:inccov}
	\mathit{h}^H(m)=\frac{1}{2}[(m+1)^{2H}+(m-1)^{2H}-2m^{2H}],
	\end{eqnarray}
	where $H$ is the Hurst index describing the dependency among the increments and integer valued $m,$ \cite{oksendal}.
\end{enumerate}

It is already known that these increments can be either positively or negatively correlated depending on the value of the Hurst parameter, $0<H<1$. In particular, an fBm with parameter $H=1/2$ corresponds to a standard Brownian motion which has independent increments. For $H < 1/2$, its increments are negatively correlated and they display short-range dependence. In contrast, for $H > 1/2$, the auto-covariance of the fBm increments is positive. Thus, the two consecutive increments tend to have the same directions, \cite{oksendal}.

The simulation of fBm is of crucial importance for its application in fields including economics, finance, engineering and hydrology. Therefore, the development of an algorithm to simulate an fBm is both theoretically and practically required. In the literature, there is a large number of simulation methods. For instance, the integral representation introduced by \cite{mandel} is used for a direct approximation of fBm. In another study,  \cite{hosking1984modeling} generates an fBm by developing the method that implicitly computes the fGn covariance matrix. Another approach is the fast Fourier transform (FFT) method developed by \cite{davie}. In this method, FFT algorithm is used in order to generate an fGn sample. Then, the covariance matrix of fGn is embedded in a circulant covariance matrix, and this circulant matrix is diagonalized with an FFT algorithm. The Cholesky method proposed by \cite{asmussen1998stochastic} relies on the Cholesky decomposition. The other approximate and fast technique is the random midpoint displacement method proposed by \cite{lau1995self}. Similar to the \cite{hosking1984modeling} method, it utilizes an approach based on computing the conditional distribution of fGn. The only difference is that the generation is based on the conditional distribution given certain points instead of all points. Another way of generating an fBm process is the wavelet based synthesis method, see \cite{Wornell90}, \cite{abry1996wavelet}. The principle idea of this method is to write
an fBm as a weighted sum of orthonormal wavelets, where the weights are samples of independent centered Gaussian processes. The fast and efficient implementation of this method is given in \cite{abry1996wavelet}. The method firstly generates the wavelet coefficients corresponding to an orthogonal basis. Afterwards, fBm is obtained via an inverse wavelet transformation. \cite{caglar2000simulation} summarizes and discusses these techniques and introduces a new algorithm using micropulses approximation to synthesize a fractional Brownian motion.

In this study, different from those reviewed above, we propose a new fBm simulation method through a discrete process, namely correlated random walk. To the best of our knowledge, the generation methods of fBm using the discrete processes go back to \cite{dasgpfrac}, which uses the independent binary random variables and the stochastic integral representation of fBm to approximate an fBm. On the other hand, \cite{sottinen2001fractional} defines a random walk, for long range dependence case which converges weakly to fBm using a kernel function that converts the standard Brownian motion to fBm. \cite{szabados2001strong} benefits from the moving average of an appropriately nested sequence of random walks uniformly converging to fBm when $H\in (\frac{1}{4},1)$. This approximation uses the discrete form of moving average representation. As another way of fBm generation by discrete processes, \cite{enriquez2004simple} proves that the normalized persistent random walk converges weakly to fBm. The construction relies on the correlated jumps in such a way that the probability of having the same jump as the previous one, which is the persistence parameter, defines the correlation of random walk. \cite{konstantopoulos2004convergence} have introduced that scaled random walk using the weighted sum of independent and identically distributed random variables converge to fBm under the sufficient condition for the weak convergence of normalized sums to fBm with $H>\frac{1}{2}$. Similar to \cite{konstantopoulos2004convergence}, \cite{lindstrom2007random} uses the same approximation for the case  $H<\frac{1}{2}$. 

In the \cite{donsker1951invariance} Theorem it is proven that standard Brownian motion can be constructed by independent increment random walk using the Central Limit Theorem. As an analogue of this idea, fBm can also be constructed by dependent increment random walk. Except for \cite{enriquez2004simple} the above mentioned studies use independent increment random walk. \cite{enriquez2004simple} construction depends on the persistent random walk with a persistence parameter corresponding to the probability of producing the same jump with the previous one. 

In this study, we propose a new algorithm, where we generate correlated binary random variates with proportions $p.$ An explicit density is assigned to the values of $p$ using the link between the correlation of multivariate Gaussian and the correlation of their dichotomization along with the relation between the correlation and the persistence parameter given in \cite{enriquez2004simple}. The correlations of increments of this random walk are obtained from the one correlation of increments of fBm for a given Hurst parameter and the marginal proportions, $p.$ We prove that the normalized sum of independent trajectories of this newly proposed correlated random walk yields a discrete Gaussian process by the Central Limit Theorem. Furthermore, its scaling limit is the desired fractional Brownian motion with $H\in [1/2,1],$ since it owns the correlation structure which satisfies the conditions specified in \cite{taqqu1975weak}.  Our newly proposed algorithm generalizes the construction given in \cite{enriquez2004simple}, Thm 1, in a way that it uses the same correlation of increments, but with the Bernoulli $p$ random variables as increments where $p$ has a certain density function, rather than using only Bernoulli $p=1/2$ random variables. This newly proposed algorithm is simple to follow and implement by a broad range of researchers working in various fields of applied sciences and it also has theoretical interest as it is the exact analogous of independent random walk converging to Brownian motion for dependent random walk converging to fractional Brownian motion.

\section{The correlated random walk construction}\label{Section2}

Different from binary variables obtained by natural means such as male or female, yes or no, success or failure, some dichotomous variables can be generated via discretization of continuous ones. The dichotomous variables, such as having a high or low income or being short or tall, can be produced by assigning a threshold value to continuous variables. Despite the loss of  information, these variables are crucial for many scientific fields and they are easy to implement and use in many areas such as psychology, criminology, biology and sociology. The correlation between the two continuous variables is usually calculated by Pearson correlation, but when the two are dicotomized by a threshold term, the correlation name changes. The tetrachoric correlation coefficient is assigned for the correlation between two dichotomous variable before discretization and after discretization. The correlation between the dichotomized variables is called phi correlation coefficient. \cite{demirtas2017anatomy} emphasize that when both variables are discretized, the magnitude of these correlations can easily be transformed into the binary case under the normality assumption and the connection between the tetrachoric correlation and the phi coefficient is known.

Assume that $Z_j$'s denote the Normally distributed variables. These are dichotomized to produce $Y_j$'s which represent binary variables. Suppose $Y_1,Y_2,...,Y_j$ are $J$ binary random variables having the mean $E[Y_j]=p_j$ for $j=1,2,....,J$ and the correlation $Corr[Y_j,Y_k]=\sigma_{jk}$ for $j=1,2,...,J-1; k=2,3...,J$.
Let $Z=(Z_1,Z_2,...,Z_J)^T$ represent the $J$-dimensional multivariate normal random variables with zero mean and $Corr[Z_j,Z_k]=\delta_{jk}$ for $j=1,2,...,J-1; k=2,3...,J$. Then, the link between the tetrachoric correlation ($\delta_{jk}$) and the phi coefficient ($\sigma_{jk}$) is as follows:
\begin{equation}
\label{eq:corbt}
\sigma_{jk} =\frac{\Phi[z(p_j),z(p_k),\delta_{jk}]-p_jp_k}{\sqrt{p_j(1-p_j)p_k(1-p_k)}}, 
\end{equation}
where  $z(p)$ represents $p$ th quantile of standard normal distribution and\\
$\Phi[z(p_j),z(p_k),\delta_{jk}]$ be a cumulative distribution function of standard bivariate normal with correlation coefficient $\delta_{jk}$ for $j=1,2,...,J-1; k=2,3...,J.$ Explicitly, 
$$\Phi[z(p_j),z(p_k),\delta_{jk}]=\int_{-\infty}^{z(p_j)}\int_{-\infty}^{z(p_k)} f(z(p_j),z(p_k),\delta_{jk}) dz(p_j)dz(p_k)$$ 
where $f(z(p_j),z(p_k),\delta_{jk})=[2\pi^{-1}(1-\delta_{jk})^{-\frac{1}{2}}]\times exp[-({z(p_j)}^2-2z(p_j)z(p_k)\delta_{jk}+{z(p_k)}^2]/2(1-\delta_{jk}^2)$ with $j=1,2,...,J-1; k=2,3...,J.$ Note that, on the condition $\sigma_{jk}$ is within the correlation range mentioned in \cite{demirtas2011practical}, the solution can be obtained uniquely. After generating the normal outcomes $(Z_j)$, binary variables $(Y_j)$ can be constructed by setting $Y_j=1$ if $Z_j\leq z(p_j)$ and $0$ if otherwise. Equivalently, by setting $Y_j=1$ if $Z_j\geq z(1-p_j)$, binary variables could be created because there will be no change in the correlation. Note also that the discretization proportion corresponds to the proportion of 1's which we observe by dichotomization. 

Our main interest in this study is to generate an fBm via correlated random walk as an analogous method of generating Brownian motion through identically independently distributed random walk. For this purpose, the discretization context explained above can be implemented to the increments of fBm, in order to create a random walk with an explicit correlation structure.
Therefore, we derive the correlated random walk by setting $Y_j=1$ if $Z_j \leq z(p_j)$ and $Y_j=-1$ if  $Z_j> z(p_j)$. This creates the increments of correlated random walk and by taking the cumulative sum of them we can construct a correlated path. Notice that, Eqn. \eqref{eq:corbt} is assured when binary variables are observed by setting $Y_j=1$ if $Z_j\leq z(p_j)$ and $0$ otherwise. Fortunately, binary variables $Y$ can also be assigned the values of $+1$ and $-1$ by taking the linear transformation and it can be checked that the link between the tetrachoric correlation and the phi-coefficient given in Eqn. \eqref{eq:corbt} remains the same. Now, for simplicity let us consider the discretization proportions $p_j$ and $p_k$ equal the same proportion $p,$ in which case the link given in Eqn. \eqref{eq:corbt} can be reduced to,
\begin{equation}
\label{eq:simplecorbt}
\sigma_{jk} =\frac{\Phi[z(p),z(p),\delta_{jk}]-p^2}{{p(1-p)}}, 
\end{equation}
where  $z(p)$ represents $p$th quantile of standard normal distribution and\\
$$\Phi[z(p),z(p),\delta_{jk}]=\int_{-\infty}^{z(p)}\int_{-\infty}^{z(p)} f(z(p),z(p),\delta_{jk}) dz(p)dz(p)$$ is the cumulative distribution function of standard bivariate normal with correlation coefficient $\delta_{jk}$ where $f(z(p),z(p),\delta_{jk})=[2\pi^{-1}(1-\delta_{jk})^{-\frac{1}{2}}]\times exp[-({z(p)}^2-2z(p)^2\delta_{jk}+{z(p)}^2]/2(1-\delta_{jk}^2)$ with $j=1,2,...,J-1; k=2,3...,J.$ 

The increment process, fGn, $\{B^{H}(n+1)-B^{H}(n): n=0,1,2,...\},$ are jointly Gaussian variates with zero mean and covariance given in Eqn. \eqref{eq:inccov}. Then, the correlation between $B^{H}(n+m+1)-B^{H}(n+m)$ and $B^{H}(n+1)-B^{H}(n)$ equals,
\begin{eqnarray}
\label{eq:inccor}
\delta^{H}_{m}=\frac{1}{2}[(m+1)^{2H}+(m-1)^{2H}-2m^{2H}],
\end{eqnarray}
with integer valued $m$ and Hurst parameter $H\in(0,1)$. 
As a consequence of these properties of fBm the one step dichotomized variates of these increments would have the following correlation, phi-coefficient,
\begin{equation}
\label{eq:corrwfrominc}
\sigma_{n,n+1} =\frac{\Phi[z(p),z(p),\delta^{H}_{1}]-p^2}{{p(1-p)}}, 
\end{equation}
for $n=0,1,2,...$ by replacing the tetrachoric correlation with the correlation of fGn. And, here in this study, in order to generate fBm via correlated random walk, we first propose to generate correlated binary variates, $\pm 1$'s where $P(Y_n=1)=p,$ $P(Y_n=-1)=1-p$ with $m-$step correlation 
\begin{equation}
\sigma_{n,n+m} =\frac{(4\Phi[z(p),z(p),\delta^{H}_{1}]-4p+1)^m-(2p-1)^2}{4p(1-p)}
\end{equation} Second, we propose to take the cumulative sum of these binary random variates to construct a correlated trajectory. In such a simulation study, we have to answer two main questions which are, 
\begin{itemize}
	\item Does this simulated trajectory converge to fBm?
    \item If it does, for which values of probability $p$ is the convergence  satisfied?
\end{itemize}

\section{The convergence of correlated random walk}\label{Section3}
In this Section, we prove the convergence of the proposed correlated random walk given in Section \ref{Section2} to fBm. \cite{enriquez2004simple} introduces the construction of fBm using random walk with certain persistence parameters. The persistent random walk is defined as a discrete time process including the jumps of size $\pm 1,$ whose probability of making the same jump as the previous one is the parameter of persistence. Let $X^{\rho}$ be a discrete process with persistence parameter, $\rho \in [0,1]$ then,
\begin{enumerate}
	\item $X^{\rho}_0=0,$ $P(X^{\rho}_1=-1)=1/2$ and $P(X^{\rho}_1=+1)=1/2,$
	\item for all $n\geq 1,$ $\epsilon^{\rho}_n=X^{\rho}_n-X^{\rho}_{n-1}$ is equal to $-1$ or $+1$ a.s.,
	\item $P(\epsilon^{\rho}_{n+1}=\epsilon^{\rho}_n|\sigma(X_{k}^{\rho},0\leq k\leq\ n))=\rho$ for all $n\geq 1.$
\end{enumerate}
Furthermore, by the help of conditioning on $\epsilon^{\rho}_1,$ it can be easily seen that for all $n\geq 1,$ we have $P(\epsilon^{\rho}_{n}=\pm 1)=1/2.$ Hence, the correlation among $n$ distance time steps is,  
\begin{equation}
\label{fixedcor}
E[\epsilon^{\rho}_{m}\epsilon^{\rho}_{m+n}]=(2\rho-1)^n
\end{equation} for all $n \geq 0,$ $m \geq 1$. In \cite{enriquez2004simple} study additional randomness into persistence  parameter $\rho$ is introduced where $P^{\rho}$ stand for the law of $X^{\rho}$ for a given random persistence. After forming a new probability measure $\mu$ on $[0,1]$, the law of persistent random walk corresponding to this $\mu$ is $P^\mu:\int_{0}^1{P^{\rho} d\mu(\rho)}$ and under $P^\mu$ for the process $X^\mu$, the correlation among $n$ distance time steps is computed by
\begin{equation}
\label{corprop}
E[\epsilon^{\mu}_{m}\epsilon^{\mu}_{m+n}]=\int_{0}^{1}(2\rho-1)^nd\mu(\rho)
\end{equation}
for all $n \geq 0$ , $m \geq 1$. Additionally, \cite{enriquez2004simple} proves that the persistent random walk $X^{\mu}$ with the law of $P^{\mu}$ weakly converges to fBm $B_{H}(t)$ by Lemma 5.1 of \cite{taqqu1975weak}. The convergence can be shown in two steps. First, the summation of great number of trajectories converge to  discrete Gaussian process by the Central Limit Theorem, and second this discrete Gaussian process  converges to fBm when rescaled as it satisfies the correlation condition stated in \cite{taqqu1975weak}. Now, let us start by exploring the connection between the persistant random walk and the correlated binary variates we have proposed in Section \ref{Section2}. 
\begin{lemma}
	\label{rho}
Let $\rho$ be the persistence parameter. Then, $$\rho=2\Phi[z(p),z(p),\delta^{H}_{1}]-2p+1,$$ where $\delta^{H}_{1}=\frac{1}{2}[2^{2H}-2].$
\end{lemma}
\prf
Let $Y_n$ denote the jump size of random walk at the $n^{th}$ step. Then  $\rho=P(Y^{\rho}_{n+1}=Y^{\rho}_n|\sigma(Y_{k}^{\rho},0\leq k\leq n))$ for all $n\geq1.$ Therefore in terms of dichotomization we have
\begin{eqnarray}
\label{lnk2}
&&P(Y^{\rho}_{n+1}=Y^{\rho}_n|\sigma(Y_{k}^{\rho},0\leq k\leq n))\nonumber\\ &=&P(Y^{\rho}_{n+1}=1,Y^{\rho}_n=1|\sigma(Y_{k}^{\rho},0\leq k\leq n))\nonumber\\
&+&P(Y^{\rho}_{n+1}=-1,Y^{\rho}_n=-1|\sigma(Y_{k}^{\rho},0\leq k\leq n))\nonumber\\
&=&P(Z^{\rho}_{n+1}\leq z(p),Z^{\rho}_{n}\leq z(p))+P(Z^{\rho}_{n+1}> z(p),Z^{\rho}_{n}> z(p))\nonumber
\end{eqnarray}
for some $Z_n$ and $Z_{n+1}$ bivariate Normal random variables with correlation coefficient $\delta _{n,n+1}.$
Recall that, $P[Z_n \leq z(p)]=p$, $P[Z_n > z(p)]= 1-p$ and
\begin{eqnarray}
\label{eq:cdf}
P[Z_n\leq z(p),Z_{n+1}\leq z(p)]=\Phi[z(p),z(p),\delta_{n,n+1}].
\end{eqnarray}
Besides,
\begin{eqnarray}
\label{eq:9}
P[Z_n > z(p),Z_{n+1} \leq z(p)]&=&P[Z_{n} \leq z(p)]-P[Z_{n} \leq z(p),Z_{n+1} \leq z(p)]\nonumber\\
&=& p -\Phi[z(p),z(p),\delta _{n,n+1}],
\end{eqnarray}
and we obtain
\begin{eqnarray}
\label{eq:lnk}
P[Z_n > z(p),Z_{n+1} > z(p)]&=&P[Z_n > z(p)]-P[Z_n > z(p),Z_{n+1} \leq z(p)]\nonumber\\
&=& (1-p)-(p -\Phi[z(p),z(p),\delta _{n,n+1}]\nonumber)\\
&=& \Phi[z(p),z(p),\delta _{n,n+1}]-2p+1.
\end{eqnarray}
Hence, by Eqns.\eqref{eq:cdf} and \eqref{eq:lnk} we have
\begin{eqnarray}
\label{eq:relationship}
\rho=2\Phi[z(p),z(p),\delta^{H}_{1}]-2p+1
\end{eqnarray}
Note that, by the arguments provided for obtaining Eqn. \eqref{eq:corrwfrominc}, we can replace $\delta_{n,n+1}$ with $\delta^{H}_{1}$ to conclude the statement.
\finish
\begin{prop}\label{nstep}
	Let $Y_n$ and $Y_{n+1}$ be the dichotomized variables of one step increments of fBm then the $n-$step correlation among them is 
	\begin{eqnarray}\label{corpm1}
	r(Y_{m},Y_{m+n})=\frac{(4\Phi[z(p),z(p),\delta^{H}_{1}]-4p+1)^n-(2p-1)^2}{4p(1-p)}
	\end{eqnarray}
\end{prop}
\prf Given $\sigma(Y_{k}^{\rho},0\leq k\leq n)$, the random variables $Y_n$ and $Y_{n+1}$ are the dichotomized variables of one step increments of fBm which are bivariate Gaussian. Therefore their correlation equals,
\begin{eqnarray}
r(Y_{n},Y_{n+1})=\frac{\Phi[z(p),z(p),\delta^{H}_{1}]-p^2}{{p(1-p)}}=\frac{E[Y_{n}Y_{n+1}]-(2p-1)^2}{4p(1-p)}
\end{eqnarray}
and hence $E[Y_nY_{n+1}]=4\Phi[z(p),z(p),\delta^{H}_{1}]-4p+1.$ On the other hand
for all $n\geq 1$ we have 
\begin{eqnarray}
E[Y_{n+1}|\sigma(Y_{k},0\leq k\leq n)]&=&E[Y_{n+1}1_{Y_{n}=1}|\sigma(Y_{k},0\leq k\leq n)]\\
&+&E[Y_{n+1}1_{Y_{n}=-1}|\sigma(Y_{k},0\leq k\leq n)]=(2\rho-1)Y_n\nonumber
\end{eqnarray} and thus $E[Y_nY_{n+1}]=2\rho-1=4\Phi[z(p),z(p),\delta^{H}_{1}]-4p+1.$ As a result, we have $\rho=2\Phi[z(p),z(p),\delta^{H}_{1}]-2p+1$ which coincides with the result obtained in Lemma \ref{rho}. Consequently, as given in Proposition 1 of \cite{enriquez2004simple}, one can show that conditioning on $\sigma(Y_{k}^{\rho},0\leq k\leq m+n)$ by induction for all $m\geq1,~n\geq 1,$ $E(Y_{m},Y_{m+n})=(2\rho-1)^n.$ Thus, by Lemma \ref{rho}, in our proposed method, we take the correlation equal to
\begin{eqnarray}
r(Y_{m},Y_{m+n})&=&\frac{E[Y_mY_{m+n}]-(2p-1)^2}{4p(1-p)}\nonumber\\
&=&\frac{(4\Phi[z(p),z(p),\delta^{H}_{1}]-4p+1)^n-(2p-1)^2}{4p(1-p)}
\end{eqnarray}


\finish

As a result of these relations, it is observed that both the persistence parameter and the correlation coefficients can be written in terms of the parameter $p$ of the marginal distributions of the binary variates, $\pm 1$'s. In fact, note that the persistence parameter is a function of $p$ and $H,$ only. Due to the properties of fBm, the Hurst parameter H is a constant in the interval $[0,1],$ which therefore introducing randomness to the persistence parameter results in introducing randomness to the parameter $p.$ 

In Theorem \ref{teo}, as a consequence of obtaining the link between the persistence parameter and the parameter of marginal proportions of the correlated binary variates, we are able to answer both questions introduced at the end of Section \ref{Section2}. We prove that the proposed correlated random walk with proportion satisfying an explicit density in fact converges to fBm with $H\in[1/2,1]$.
\begin{theorem}\label{teo}
For $H\in [\frac{1}{2},1]$ let $p$, the marginal parameter of correlated binary random variables $\pm 1,$ have the density
\begin{eqnarray}\label{density}
g(p)&=&(1-H) 2^{3-2 H}\left(1-\frac{\left(\frac{\left(4 \Phi\left[z(p), z(p), \delta_{1}^{H}\right]-4 p+1\right)^{n}-(2 p-1)^{2}}{4 p(1-p)}\right)^{1 / n}+1}{2}\right)^{1-2 H}\nonumber\\
&&\frac{d(\frac{\left(\frac{\left(4 \Phi\left[z(p), z(p), \delta_{1}^{H}\right]-4 p+1\right)^{n}-(2 p-1)^{2}}{4 p(1-p)}\right)^{1 / n}+1}{2})}{dp}
\end{eqnarray}
  for values of $p$ in the range $
  0 \leq \frac{\left(4 \Phi\left[z(p), z(p), \delta_{1}^{H}\right]-4 p+1\right)^{n}-(2 p-1)^{2}}{4 p(1-p)} \leq 1
  $ and $(X^{H,i})_{i\geq 1}$ be a sequence of standardized independent process of this law then, $$\lim_{N\to\infty}\lim_{M\to\infty} a_H\frac{X_{[Nt]}^{H,1}+X_{[Nt]}^{H,2}...+X_{[Nt]}^{H,M}-M(Nt)(2p-1)}{N^H\sqrt{M4p(1-p)}}=B^{H}(t)$$ where $a_{H}=\sqrt{\frac{H(2H-1)}{\Gamma{(3-2H)}}},$  N is the number of time steps and M be the number of trajectories, $B^H(t)$ is the fBm with Hurst index $H\in[1/2,1].$
\end{theorem}
\prf
In order for a persistent random walk to converge to fBm, the density imposed on the persistence parameter$\rho\in[\frac{1}{2},1]$ is given as;
\begin{equation}
\label{ro}
f({\rho})=(1-H)2^{3-2H}(1-\rho)^{1-2H}
\end{equation}
for values of $H\in [\frac{1}{2},1],$ see \cite{enriquez2004simple}.
Now, let $Y_k^{H,i}$ be $k$ th step of $i$ th trajectory consisting of the correlated binary variables ($\pm 1$) with marginal proportions $p$ and $1-p$ respectively for $k\geq1$ and $1\leq i\leq M$ where $M$ represents the number of trajectories. Now, let $X_k^{H,i}=X_{k-1}^{H,i}+Y_{k}^{H,i}, k\geq 1$ with $X_0=0.$ Note that, $E(Y_{k}^{H,i})=2p-1,$  for $k\geq 1$ and so $E(X_k^{H,i})=k(2p-1).$ Also for all $k\geq 1$, 
\begin{eqnarray}
E((X^{H,i}_{k})^2)&=&E((X_{k-1}^{H,i}+Y_{k+1}^{H,i})^2)=E(E(X_{k-1}^{H,i}+Y_{k+1}^{H,i})^2|Y_{k+1}^{H,i}))\nonumber\\
&=&pE((X_{k-1}^{H,i}+1)^2)+(1-p)E((X_{k-1}^{H,i}-1)^2)\nonumber\\
&=&E((X_{k-1}^{H,i})^2)+2(k-1)(2p-1)^2+1
\end{eqnarray}
By induction we have
$E((X^{H,i}_{k})^2)=2(2p-1)^{2}\sum_{j=2}^{k}(j-1)+k=k+k(k-1)(2p-1)^{2}$ for all $k\geq 1.$ Hence, $V(X^{H,i}_{k})=k(1-(2p-1)^2)$ for all $k\geq 1.$

Then, by the central limit theorem
$$\lim_{M\to\infty}{\frac{X_k^{H,1}+X_k^{H,2}...+X_k^{H,M}-Mk(2p-1)}{\sqrt{M}\sqrt{4p(1-p)}}}=Z^H_k$$ is a discrete centered Gaussian process for $k\geq 1$ with $E[Z^H_k]=0$ and $V[Z^H_k]=\frac{V[X_k^H]}{4p(1-p)}$ since it is the sum of $M$ independent identically distributed random variables.  Now for $k\geq 1$, let us define the increment process, 
\begin{eqnarray}
G^H_k:&=&(Z^H_{k+1}-Z^H_k)=\lim_{M\to\infty}\frac{\sum_{i=1}^{M}(X_{k+1}^{H,i}-X_k^{H,i})-M(2p-1)}{\sqrt{M4p(1-p)}}\nonumber\\
&=&\lim_{M\to\infty}\frac{\sum_{i=1}^{M}Y_{k+1}^{H,i}-M(2p-1)}{\sqrt{M4p(1-p)}}\nonumber
\end{eqnarray}
which is the sum of independent, identically distributed random variables. Therefore, by the central limit theorem, it is Gaussian and stationary with $E[G^H_k]=0$ and $E[(G^H_k)^2]=1$ for $k\geq1.$ The n step correlation is;
\begin{eqnarray}
r(n)&=&E[G^H_kG^H_{k+n}]=r(Y_{k+1}^{H},Y_{k+n+1}^{H})=\sigma_{k,k+n}
\end{eqnarray}
Now, the $n$ step correlation of $\pm1'$s, $\sigma_{k,k+n}$ is 
\begin{eqnarray} \sigma_{k, k+n} &=&\frac{E\left[Y_{k} Y_{k+n}\right]-(2 p-1)^{2}}{4 p(1-p)}=\frac{(2 \rho-1)^{n}-(2 p-1)^{2}}{4 p(1-p)}\nonumber\\ &=&\frac{\left(4 \Phi\left[z(p), z(p), \delta_{1}^{H}\right]-4 p+1\right)^{n}-(2 p-1)^{2}}{4 p(1-p)} \end{eqnarray}
and a distribution can be imposed to $p$ by the help of the distribution of $\rho$ given in Eqn \eqref{ro} for which the convergence is satisfied. 
Thus using the transformation of random variables on the density of $\rho,$ the density of $p$ is obtained as
\begin{eqnarray}
g(p)&=&(1-H) 2^{3-2 H}\left(1-\frac{\left(\frac{\left(4 \Phi\left[z(p), z(p), \delta_{1}^{H}\right]-4 p+1\right)^{n}-(2 p-1)^{2}}{4 p(1-p)}\right)^{1 / n}+1}{2}\right)^{1-2 H}\nonumber\\
&&\frac{d(\frac{\left(\frac{\left(4 \Phi\left[z(p), z(p), \delta_{1}^{H}\right]-4 p+1\right)^{n}-(2 p-1)^{2}}{4 p(1-p)}\right)^{1 / n}+1}{2})}{dp}
\end{eqnarray}
As a consequence, we observe that 
\begin{eqnarray}
&&r(n)=E[G^H_kG^H_{k+n}]\nonumber\\
&=&\int_{p_l}^{p_u}\resizebox{\hsize}{!}{$\frac{\left(4 \Phi\left[z(p), z(p), \delta_{1}^{H}\right]-4 p+1\right)^{n}-(2 p-1)^{2}}{4 p(1-p)}(1-H) 2^{3-2 H}\left(1-\frac{\left(\frac{\left(4 \Phi\left[z(p), z(p), \delta_{1}^{H}\right]-4 p+1\right)^{n}-(2 p-1)^{2}}{4 p(1-p)}\right)^{1 / n}+1}{2}\right)^{1-2 H}\frac{d(\frac{\left(\frac{\left(4 \Phi\left[z(p), z(p), \delta_{1}^{H}\right]-4 p+1\right)^{n}-(2 p-1)^{2}}{4 p(1-p)}\right)^{1 / n}+1}{2})}{dp}dp$}\nonumber\\
&=&\int_{1/2}^{1}(1-H)2^{3-2H}(2v-1)^n(1-v)^{1-2H}dv\nonumber\\
&=&(2-2H)\frac{\Gamma(n+1),\Gamma(2-2H)}{\Gamma(n+3-2H)}\nonumber\\
&\underset{n\to\infty}{\sim}&{\frac{1}{a^2_H}\frac{H(2H-1)}{n^{2-2H}}}\nonumber
\end{eqnarray}
where $v=\frac{\left(\frac{\left(4 \Phi\left[z(p), z(p), \delta_{1}^{H}\right]-4 p+1\right)^{n}-(2 p-1)^{2}}{4 p(1-p)}\right)^{1 / n}+1}{2}$ and $p_l$ and $p_u$ are the lower and upper values of $p$ which satisfy the inequality
$$
0 \leq \frac{\left(4 \Phi\left[z(p), z(p), \delta_{1}^{H}\right]-4 p+1\right)^{n}-(2 p-1)^{2}}{4 p(1-p)} \leq 1
$$
 and $a^2_H$ satisfies 
$$E[a^2_H(G^H_1+...+G_{N}^H)]\underset{N\rightarrow\infty}{\sim}N^{2H}$$ as given in the proof of Thm 1 of \cite{enriquez2004simple}. Hence, by the \cite{taqqu1975weak}, Lemma 5.1 we conclude that the sum of the Bernoulli($p$) random variables stated in this Theorem converge to fBm with $H\in[1/2,1]$
\finish
\section 
{A simple algorithm for generating fractional Brownian Motion using correlated random walk}\label{sec4}
The aim of this article is to propose an alternative yet simple algorithm which uses multivariate Bernoulli random variables to generate fBm through correlated random walk. In this paper, we mainly construct a random walk with the $n-$step correlation given in Proposition \ref{nstep} and with a marginal parameter $p,$ satisfying an explicit density given in Theorem 1. We have proved that this correlated random walk converges weakly to fractional Brownian motion using the relation between $p$ and the persistence paramater $\rho$. This newly proposed algorithm generalizes the algorithm given in \cite{enriquez2004simple} by carrying the information in the persistence parameter to the marginal parameter $p$ of the correlated Bernoulli random variables. It generalizes the construction given in \cite{enriquez2004simple} since it uses the same persistence values but with the Bernoulli $p$ random variables as increments where $p$ has a certain density function, rather than using only Bernoulli $p=1/2$ random variables.  The algorithm is proposed in a simple way to be followed and implemented by various groups of researchers working in different fields of applied science. This newly proposed algorithm has also theoretical interest since it is the exact analogous method of independent random walk converging to Brownian motion. In addition, in this algorithm one makes use of the correlation of the one step increments of fBm while generating correlated Bernoulli random variables.

Initially, we obtain the values of proportion $p$ which satisfy the density given in Eqn.\eqref{density}. 
Notice that,
\begin{eqnarray}
F(a)&=&P(p \leq a) = u,\text{ where u is Uniform $(0,1)$.}\nonumber\\
&=& \int_{p_l}^{a}g_{p}(p) dp \nonumber\\
&=&=\int_{\frac{1}{2}}^{\frac{\left(\frac{\left(4 \Phi\left[z(p), z(p), \delta_{1}^{H}\right]-4 p+1\right)^{n}-(2 p-1)^{2}}{4 p(1-p)}\right)^{1 / n}+1}{2}} 2^{3-2 H}(1-H)(1-v)^{1-2H} d v \nonumber\\
&=&(H-1) 2^{3-2 H}\left.\frac{(1-v)^{2-2 H}}{2-2 H}\right|_{\frac{1}{2}} ^{\frac{\left(\frac{\left(4 \Phi\left[z(p), z(p), \delta_{1}^{H}\right]-4 p+1\right)^{n}-(2 p-1)^{2}}{4 p(1-p)}\right)^{1 / n}+1}{2}}\nonumber\\
&=&1-\left[2\left(1-\frac{\left(\frac{\left(4 \Phi\left[z(a), z(a), \delta_{1}^{H}\right]-4 a+1\right)^{n}-(2 a-1)^{2}}{4 a(1-a)}\right)^{1 / n}+1}{2}\right)\right]^{2-2 H}
\end{eqnarray}
where $v=\frac{\left(\frac{\left(4 \Phi\left[z(p), z(p), \delta_{1}^{H}\right]-4 p+1\right)^{n}-(2 p-1)^{2}}{4 p(1-p)}\right)^{1 / n}+1}{2}.$
Equivalently,
\begin{eqnarray}
\label{persd}
\frac{\left(4 \Phi\left[z(a), z(a), \delta_{1}^{H}\right]-4 a+1\right)-(2 a-1)^{2}}{4 a(1-a)}=\left(1-(1-u)^{\frac{1}{2-2 H}}\right)
\end{eqnarray} for $n=1.$
Therefore, the marginal parameter of the correlated Bernoulli random variables $p$ takes on the values which satisfy the Equation \eqref{persd}.

In the algorithm, by generating Uniform(0,1) random variable, $u,$ we first obtain the parameter $p$ of correlated Bernoulli random variables which ensure the equality \eqref{persd}. This implies that $$P(\xi(i)=\xi(i-1)|\sigma(\xi_{i},1\leq i\leq n-1))=2 \Phi\left[z(p), z(p), \delta_{1}^{H}\right]-2p+1$$ as given in Lemma \ref{rho} and Proposition \ref{nstep}. Now, consider $\xi(i-1)$ which is taken to be Bernoulli(p) random variable then note that \begin{eqnarray}\xi(i)&=&\xi(i-1)*Ber(2 \Phi\left[z(p), z(p), \delta_{1}^{H}\right]-2p+1)\nonumber\\
&+&(1-Ber(2 \Phi\left[z(p), z(p), \delta_{1}^{H}\right]-2p+1))*Ber(p)\end{eqnarray} have  Bernoulli(p) distribution for all $i=1,...,n$ and $P(\xi(i)=\xi(i-1)|\sigma(\xi_{i},1\leq i\leq n-1))$ equals $2 \Phi\left[z(p), z(p), \delta_{1}^{H}\right]-2p+1$ Hence, the $n$ step correlations among them satisfy
$\frac{\left(4 \Phi\left[z(p), z(p), \delta_{1}^{H}\right]-4 p+1\right)^{n}-(2 p-1)^{2}}{4 p(1-p)}$ as requested. Converting these Bernoulli random variables to $\pm 1$s would not change the correlation coefficient among them. Thus, after having the Bernoulli(p) $\pm1$ random variables with the desired correlation coefficients, the levels of the correlated walk is acquired by taking the summation of these increments. Finally, we use Theorem \ref{teo} which suggests that if the sums of these correlated random walk at each step which are scaled by the terms $M(Nt(2p-1))$ and $N^H\sqrt{M4p(1-p)}$ and multiplied with the constant $a_H$, the resulting process converges weakly to fBm as the number of trajectories $M$, and the number of steps $N$ become larger. 
Hence, to generate a trajectory of fBm, the steps to follow in this newly proposed algorithm are;
\begin{itemize}
\item Generate the value of $p$ from the distribution given in Theorem 1, that is find a value of $p$ which satisfy the equality \eqref{persd}.
\item Generate $$\xi(1)=Ber(p)$$ and \begin{eqnarray}\xi(i)&=&\xi(i-1)*Ber(2 \Phi\left[z(p), z(p), \delta_{1}^{H}\right]-2p+1)\nonumber\\
&+&(1-Ber(2 \Phi\left[z(p), z(p), \delta_{1}^{H}\right]-2p+1))*Ber(p)~~~ i=2,...,N.\nonumber\end{eqnarray}
\item Observe $\pm1$ using $$X(i)=2*\xi(i)-1~~~ i=2,...,N.$$
\item Repeat the first two steps $M$ times and observe $MXN$ multivariate Bernoulli ($\pm$1) random  variables where the columns are generated with the marginal parameter $p,$ and the $n-$step correlation among the columns equals  Proposition \ref{nstep}, \eqref{corpm1}.
\item First take the cumulative sum of each row and form new columns using these cumulative sums. And then take the sum of resulting columns, subtract $M(Nt)(2p-1)$ from each column sum and divide by $N^{H}\sqrt{M4p(1-p)},$ and multiply the result by $a_{H}=\sqrt{\frac{H(2H-1)}{\Gamma{(3-2H)}}}$, for large values of $M$ and $N.$
\end{itemize}

\begin{figure}[h!]
		\caption{Realizations of fBm using 100000 steps and 1000 paths}
		\label{fig:fbmpath0.8}
		\centering
		\begin{subfigure}[b]{0.90\textwidth}            
			\includegraphics[height=0.3\textwidth,width=\textwidth]{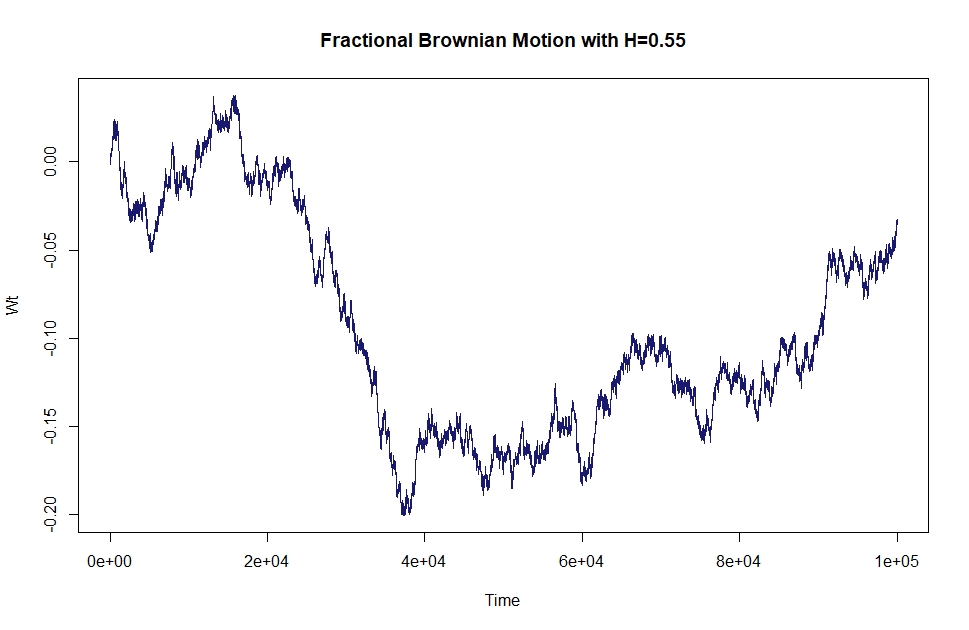}
			\caption{\cite{veitch1999wavelet} 'Discrete second derivative (DSOD)' Hurst index estimator equals 0.5426 and 'Wavelet version of DSOD' is 0.5430.}
		\end{subfigure}%
		
		\begin{subfigure}[b]{0.9\textwidth}
			\includegraphics[height=0.3\textwidth,width=\textwidth]{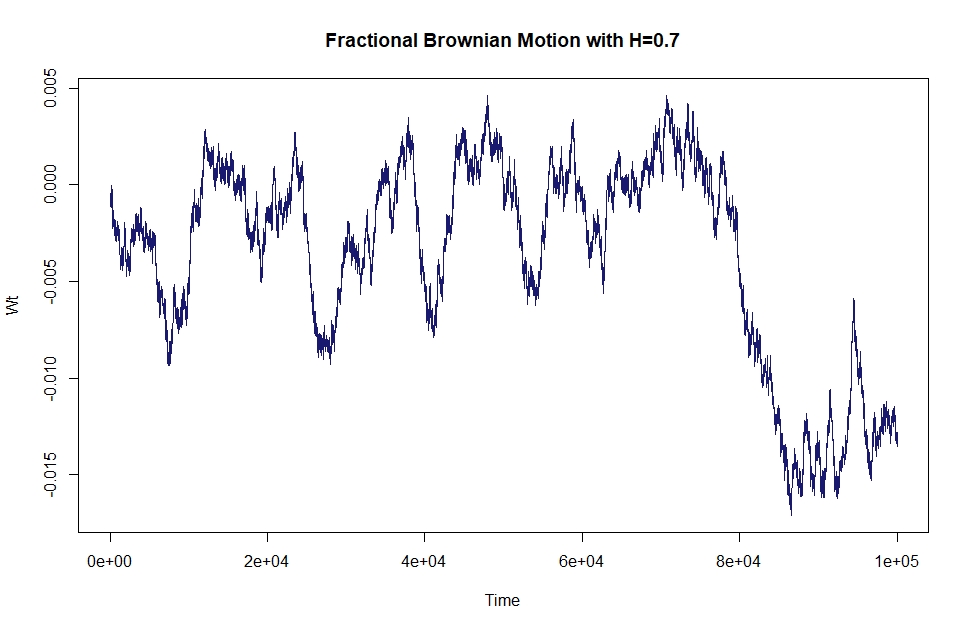}
		\caption{\cite{veitch1999wavelet} 'Discrete second derivative (DSOD)' Hurst index estimator equals 0.6960 and 'Wavelet version of DSOD' is 0.6976}
		\end{subfigure}
	
			\begin{subfigure}[b]{0.9\textwidth}
				\includegraphics[height=0.3\textwidth,width=\textwidth]{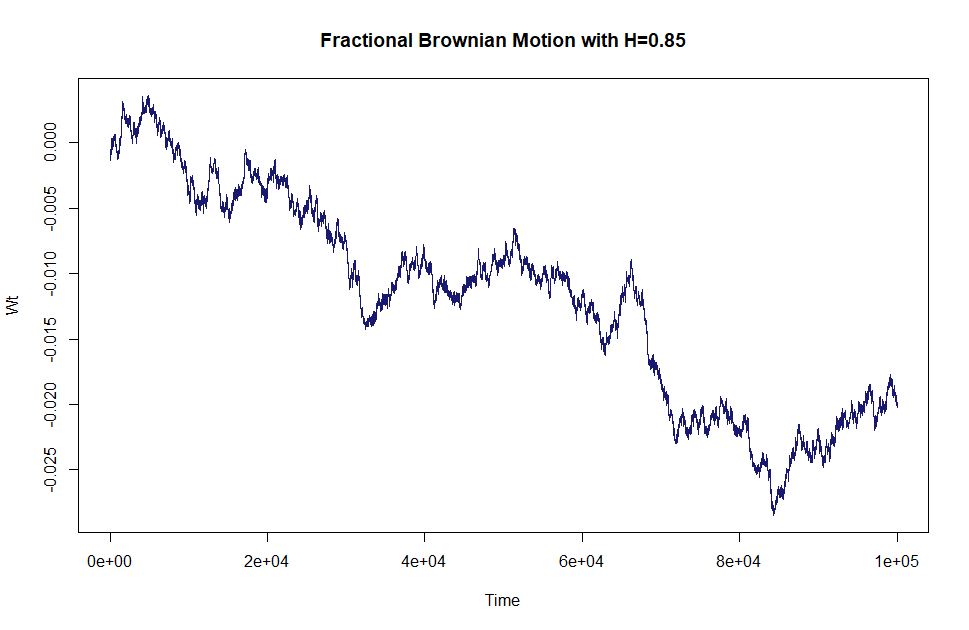}
	\caption{\cite{veitch1999wavelet} 'Discrete second derivative (DSOD)' Hurst index estimator equals 0.8429 and 'Wavelet version of DSOD' is 0.8622.}
		\end{subfigure}		
	\end{figure}

The advantage of this algorithm is that, it is based on the construction which is done analogously to the construction of the standard Brownian motion generated by independent random walk. Therefore, it provides theoretical attraction. In addition, the correlated random walk proposed here still carries the information of one step correlation of the increments of fBm. The algorithm is easier to follow for all the researchers with only the knowledge of first year Statistics courses, which makes it easy to implement. It transforms the information of the persistence parameter to the marginal parameter of correlated Bernoulli random variables and instead of generating only Bernoulli($1/2$) with persistence parameter, in this algorithm we generate Bernoulli(p) random variables with a certain distribution assigned on $p.$ Hence this method generalizes the algorithm proposed in \cite{enriquez2004simple}, in the way that it uses the same persistence of increments of fBm but with Bernoulli(p) random variables with certain distribution assigned on $p$ rather than only using the Bernoulli($1/2$) variables. The number of computations and the memory required by this algorithm is same as \cite{enriquez2004simple} algorithm. That is, for an $N$ step trajectory the number of computations is a power of $N$ between $1$ and $2$ and the memory required by the computations is a power of $N$ between $0$ and $1,$ which makes the algorithm efficient compared to the other suggested methods. Comparisons can be found in \cite{enriquez2004simple} study. The proposed algorithm is illustrated by Figure \ref{fig:fbmpath0.8} with Hurst parameters, 0.55, 0.7 and 0.85 respectively. For each realization the Hurst parameters are estimated using \cite{veitch1999wavelet} method and the results are presented. 
In fact, the convergence is satisfied for less number of steps however the \cite{veitch1999wavelet} method requires at least 10000 steps. For this reason, we have taken number of steps as 100000 in the above illustrations. We are also presenting a realization and its estimator for 10000 steps and 10000 paths in the following Figure \ref{fig:fbmpath0.7}.
	\begin{figure}[h!]
\centering
	\includegraphics[height=0.6\textwidth,width=0.8\textwidth]{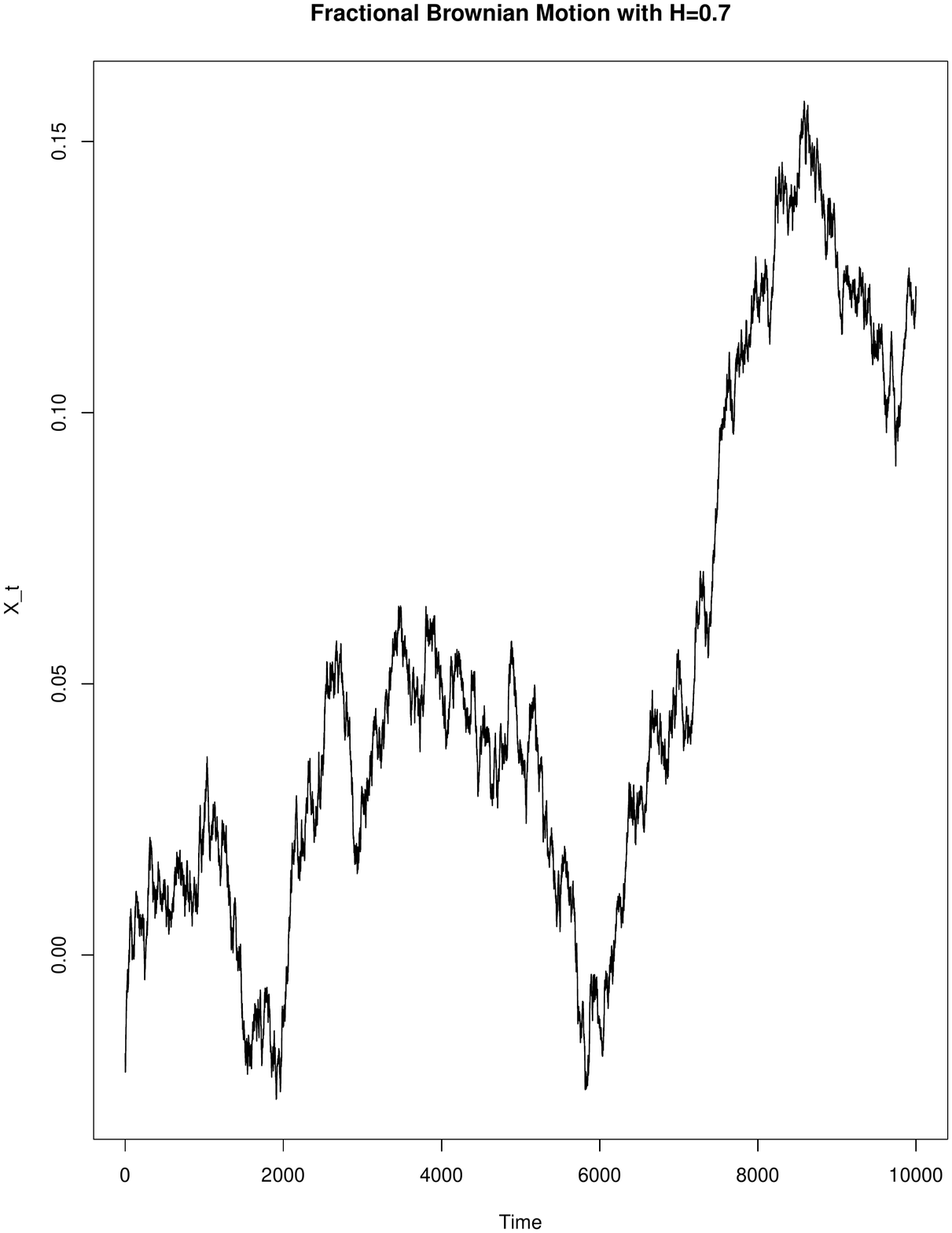}
	\caption{A realization of fBm with H=0.7 using 10000 steps and 10000 paths. The \cite{veitch1999wavelet} 'Discrete second derivative (DSOD)' Hurst index estimator equals 0.6840 and 'Wavelet version of DSOD' is  0.6879. }
	\label{fig:fbmpath0.7}
	\end{figure}

\begin{figure}[h!]
	\centering
	\includegraphics[height=0.6\textwidth,width=0.8\textwidth]{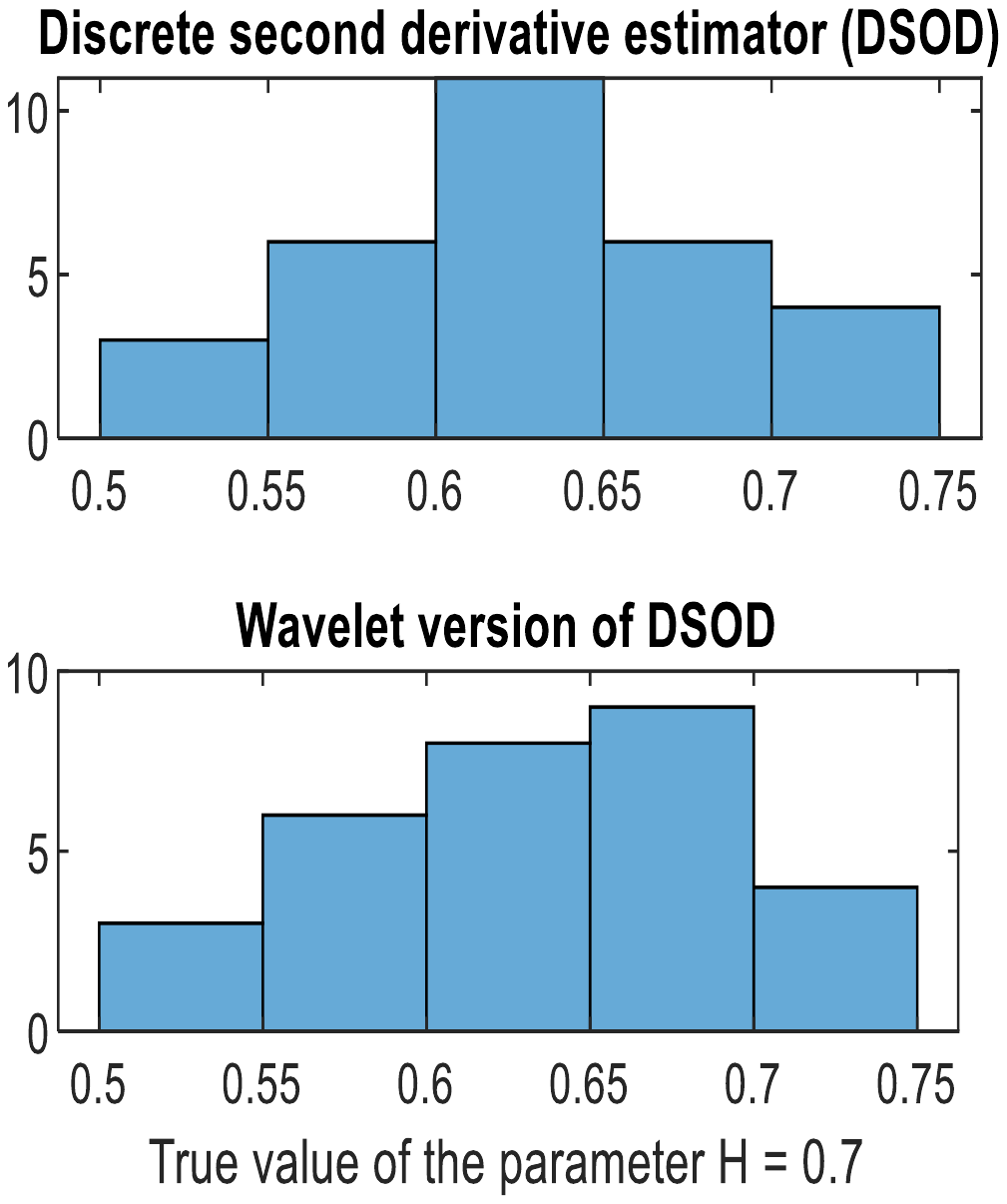}
	\caption{30 estimators of Hurst parameter of fBm with H=0.7 using 10000 steps and 10000 paths.}
	\label{fig:fbmhist0.7}
\end{figure}

We have also estimated the Hurst parameter of 30 realizations of our proposed algorithm using \cite{veitch1999wavelet} Hurst index estimation method. The histogram of these estimators are illustrated in Figure \ref{fig:fbmhist0.7} for realizations with H=0.7, 10000 steps, 10000 paths.  




\bibliographystyle{apalike}
\renewcommand\bibname{References}
\bibliography{Gencorrw}

\end{document}